\documentclass{iau}
\usepackage{graphicx}

\usepackage{epsfig,subfigure,amsmath}
\usepackage{color}
\usepackage{hyperref}
\usepackage{natbib}

\definecolor{purple}{rgb}{1,0,1}

\newcommand{\hmpc}{$h^{-1}$Mpc}

\newcommand{\beq}{\begin{equation}}
\newcommand{\eeq}{\end{equation}}
\def\vide{\textsc{vide}}

\def\ap{Alcock-Paczy\'{n}ski}

\def\apjl{ApJ}

\def\apjs{ Astrophys. J. Supp.}

\def\aap{Astron. \& Astrophys.}

\title[Answers from the Void]{Answers from the Void: VIDE and its Applications}

\author[P.M. Sutter, N. Hamaus, A. Pisani, and B.D. Wandelt]   
{P.M. Sutter$^{1,2,3}$\thanks{email: {\tt sutter@iap.fr}},
N. Hamaus$^{1,2}$,
A. Pisani$^{1,2}$, 
G. Lavaux$^{1,2}$, and
B.D. Wandelt$^{1,2,4,5}$}

\affiliation{
$^1$Sorbonne Universit\'{e}s, UPMC Univ Paris 06, UMR7095, F-75014, Paris, France \\ 
$^2$CNRS, UMR7095, Institut d'Astrophysique de Paris, F-75014, Paris, France \\
$^3$Center for Cosmology and AstroParticle Physics, Ohio State University, Columbus, USA \\
$^4$Department of Physics, University of Illinois at Urbana-Champaign, Urbana, USA \\
$^5$Department of Astronomy, University of Illinois at Urbana-Champaign, Urbana, USA \\
}

\pubyear{2014}
\volume{308}  
\pagerange{}
\jname{The Zeldovich Universe: Genesis and Growth of the Cosmic Web}
\editors{Rien van de Weygaert, Sergei Shandarin, Enn Saar, and Jan Einasto}

\begin{document}

\maketitle

\begin{abstract}
We discuss various applications of
\vide, the Void IDentification and Examination toolkit, an
open-source Python/C++ code for finding cosmic voids in galaxy redshift surveys 
and $N$-body simulations.
Based on a substantially enhanced version of \textsc{ZOBOV}, \vide~not only finds voids, but also
summarizes their properties, extracts statistical 
information, and provides
a Python-based platform for more detailed analysis, such as
manipulating void catalogs and particle members, filtering, plotting, 
computing clustering statistics, stacking, comparing catalogs, and 
fitting density profiles.
\vide~also provides significant additional functionality for 
pre-processing inputs: for example, \vide~can work with volume- or 
magnitude-limited galaxy samples with arbitrary survey geometries, 
or dark 
matter particles or halo catalogs in a variety of common formats.
It can also randomly subsample inputs
and includes a Halo Occupation Distribution model for 
constructing mock galaxy populations.
\vide~has been used for a wide variety of applications, from 
discovering a universal density profile to estimating 
primordial magnetic fields, and
is publicly available at  
{http://bitbucket.org/cosmicvoids/vide\_public}
and
\mbox{http://www.cosmicvoids.net}.
\end{abstract}


\section{Introduction}

First discovered over thirty years ago~\citep[e.g.,][]{Gregory1978}, 
cosmic voids are emerging as a novel and innovative probe of both 
cosmology and astrophysics through their
intrinsic properties~\citep[e.g.,][]{Bos2012}, 
exploitation of their statistical isotropy via the 
\ap~test~\citep{Sutter2014b}, 
or cross-correlation with the cosmic microwave 
background \citep{Ilic2013, Planck2013b, Cai2014}.
Additionally, fifth forces and modified gravity are unscreened 
in void environments, making them singular probes 
of exotic 
physics~\citep[e.g.,][]{Li2012,Clampitt2013,Carlesi2013a}.

Voids are also fascinating objects to study by themselves. 
For example, there appears to be 
a self-similar relationship between voids in dark 
matter distributions and voids in galaxies~\citep{Sutter2013a} 
via a universal density 
profile~\citep[]{Hamaus2014} (hereafter HSW).
Observationally, the anti-lensing shear 
signal~\citep{Melchior2014,Clampitt2014} and
Ly-alpha absorption measurements~\citep{Tejos2012} have illuminated the 
dark matter properties in voids.

However, there are only a few public catalogs of voids identified 
in galaxy redshift surveys, primarily the 
SDSS~\citep[e.g.,][]{Pan2011, Sutter2012a}.
And while there are many published methods for finding voids based 
on a variety of techniques, most codes remain private. 
In order to accelerate the application of voids, it is essential 
to provide easy-to-use, flexible, and strongly supported 
void-finding codes.

In this work we discuss \vide,
for Void IDentification and Examination, 
a toolkit based on the publicly-available 
watershed code \textsc{zobov}~\citep{Neyrinck2008} for finding voids 
but considerably enhanced and extended to handle a variety of simulations 
and observations~\citep{Sutter2014d}. 
\vide~also provides an extensive interface for 
analyzing void catalogs. 

\section{Void Finding}
\label{sec:finding}

\vide~is able to identify voids in $N$-body simulation snapshots produced by
\textsc{Gadget}~\citep{Gadget},
\textsc{FLASH}~\citep{Dubey2008}, \textsc{RAMSES}~\citep{ramses} 
\textsc{2HOT}~\citep{warren2013}, and generic ASCII 
files.
\vide~can either find voids directly in the full simulation, 
or randomly subsampled populations, as well as halo catalogs and 
mock galaxy samples using the Halo Occupation Distribution
formalism~\citep{Berlind2002}.
For observational datasets, to find voids in galaxy surveys
the user must provide a list of galaxy positions and a
pixelization of the survey mask
using \textsc{healpix} \citep{Gorski2005}\footnote{http://healpix.jpl.nasa.gov}.

The core of our void finding algorithm is
\textsc{zobov}~\citep{Neyrinck2008}, 
which creates a Voronoi tessellation of the tracer particle
population and uses the watershed transform to group Voronoi
cells into zones and subsequently voids~\citep{Platen2007}.
The Voronoi tessellation provides a density field estimator 
based on the underlying particle positions.

\textsc{zobov} first groups nearby Voronoi
cells into {\emph zones}, then merges adjacent zones into voids 
by finding minimum-density barriers between them.
We impose a density-based threshold within \textsc{zobov} where
adjacent zones are only added to a void
if the density of the wall between them is
less than $0.2$ times the mean particle density $\bar{n}$.
This density criterion 
prevents voids from expanding deeply into overdense structures and
limits the depth of the void hierarchy~\citep{Neyrinck2008}.

\begin{figure}
  \centering
  {\includegraphics[type=png,ext=.png,read=.png,width=0.49\columnwidth]{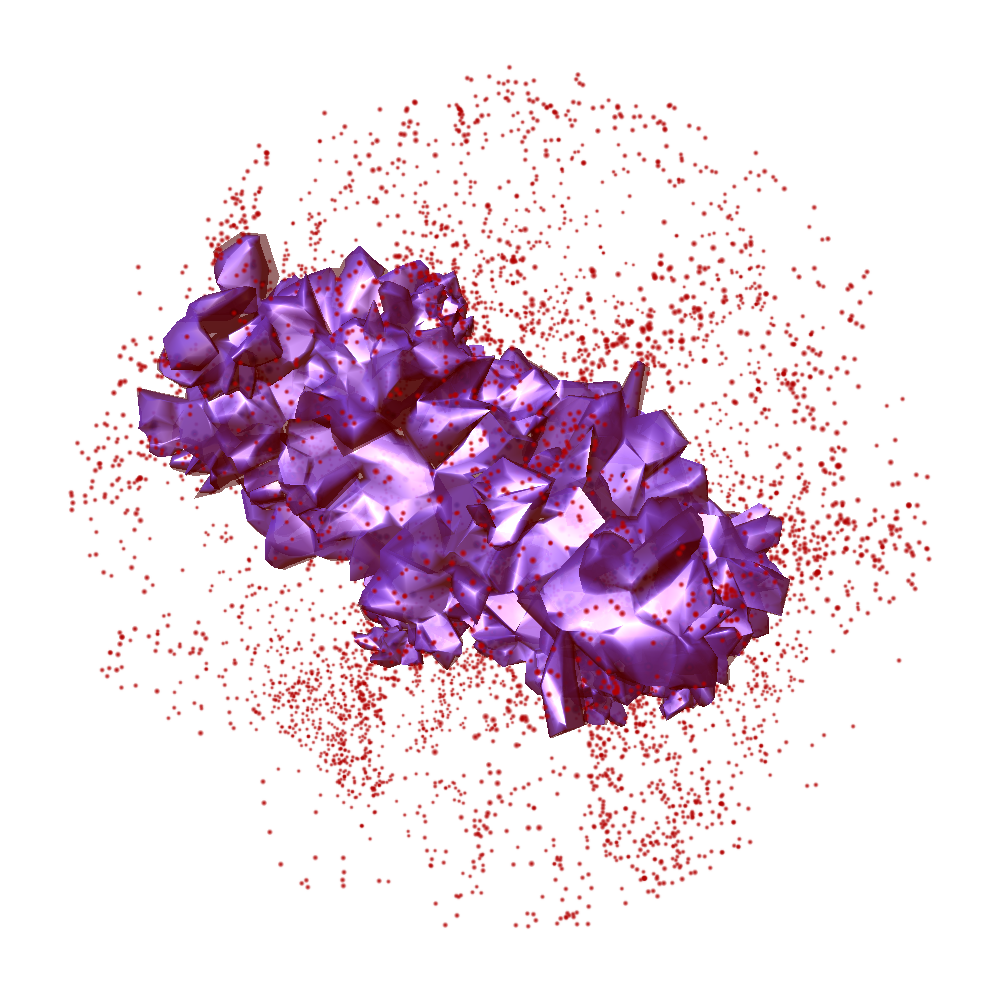}}
  {\includegraphics[type=pdf,ext=.pdf,read=.pdf,width=0.49\columnwidth]{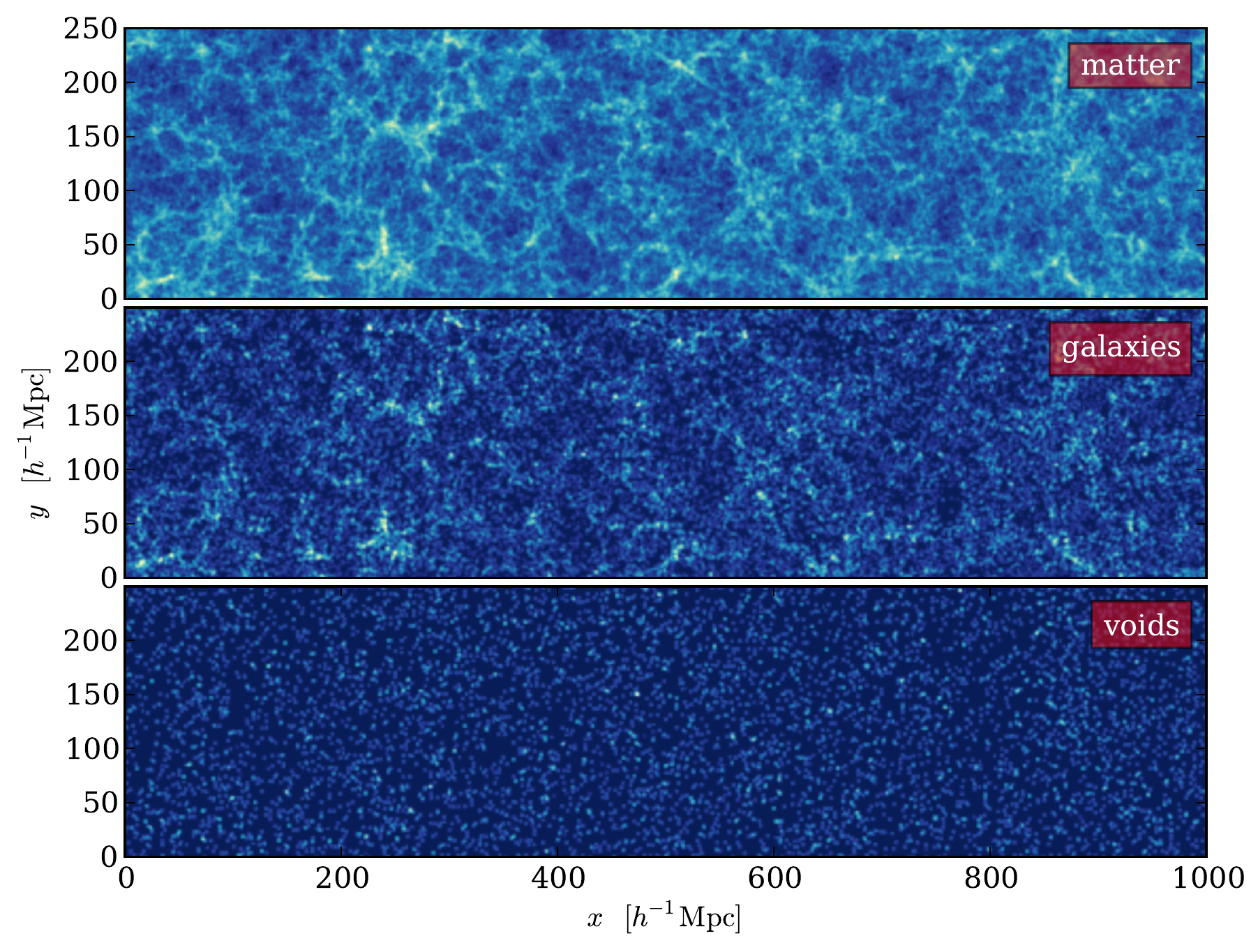}}
  \caption{Watershed void finding in \vide. 
           The left panel shows the Voronoi cells belonging to a 
           void embedded in an observed galaxy population.
           \emph{Reproduced from~\citet{Sutter2012a}}.
           The right panel shows projections of dark matter, 
           mock galaxies, and void positions in a simulation.
           \emph{Reproduced from~\citet{Hamaus2013}}.
  }
  \label{fig:void}
\end{figure}

In this picture, a void is simply a basin in the Voronoi density field 
bounded by a common set of higher-density ridgelines,
as demonstrated by Figure~\ref{fig:void}, which shows a typical 
20 \hmpc~void identified in the SDSS DR7 galaxy survey~\citep{Sutter2012a}.
This also means that voids may have any \emph{mean} density, since 
the watershed includes in the void definition all wall particles 
all the way up to the very highest-density separating ridgeline.

We have made several modifications and improvements to the original
\textsc{zobov} algorithm. First, we have strengthened the algorithm 
with respect to numerical precision by
rewriting large portions in a 
templated C++ framework.
We enforce bijectivity in the Voronoi graph, so that the tessellation 
is self-consistent, and
use an improved volume-splitting technique to minimize the number of 
difference operators at the edge of the box when
joining subregions.
Finally, we have optimized several portions of the central 
watershed algorithm, enabling the identification of voids in simulations 
with up to $1024^3$ particles in $\sim10$ hours using 
16 cores.

To prevent the growth of voids outside survey boundaries in observational 
datasets,
we place a large number of mock particles
along any identified edge and along 
the redshift caps. 
We assign essentially infinite density to these mock particles, 
preventing the watershed from including zones external 
to the survey.
Since the local volumes of the edge galaxies are arbitrary, 
we prevent these mock particles from participating in any 
void by disconnecting their adjacency links in the Voronoi graph.

We use the mean particle spacing to set a lower size limit for voids
because of Poisson shot noise. \vide~does not include any void with 
effective radius smaller than $\bar{n}^{-1/3}$, where $\bar{n}$ is the 
mean number density of the sample.
We define the effective radius as the radius of a sphere with 
the same total volume of all the Voronoi cells that make up the void.

Figure~\ref{fig:void} shows an example void population with
\vide, taken from 
the analysis of~\citet{Hamaus2013}. In this figure we show a slice 
from an $N$-body simulation, a set of mock galaxies painted onto the 
simulation using the HOD formalism discussed above, and the distribution 
of voids identified in the mock galaxies.

\vide~automatically provides some basic derived void information, such as 
the \emph{macrocenter}, or volume-weighted
center of all the Voronoi cells in the void, and the ellipticity.

\section{Post-Processing \& Analysis}
\label{sec:analysis}

\vide~provides a Python-based application programming interface (API) 
for loading and manipulating the void catalog and performing 
analysis and plotting.

Via the API the user has immediate access to all basic and derived void properties
(ID, macrocenter, radius, density contrast, RA, Dec, hierarchy 
level, ellipticity, etc.)
as well as the positions ($x$, $y$, $z$, RA, Dec, redshift),
velocities (if known), and local Voronoi volumes of all void member particles. 
The user can also access all particle and galaxy sample 
information, such as redshift extents, the mask, simulation extents 
and cosmological parameters.
Upon request the user can also load all particles in the simulation 
or observation. 

\vide~includes several built-in plotting routines.
First, users may plot cumulative number functions of multiple 
catalogs on a logarithmic scale. 
Volume normalizations are handled automatically, 
and 1$\sigma$ Poisson uncertainties are shown as shaded regions, 
as demonstrated in Figure~\ref{fig:numberfunc}. 
Secondly, users may plot a slice from a single void and its 
surrounding environment.
In these plots we bin the background particles onto a 
two-dimensional grid and plot the density on a logarithmic scale. 
We draw the void member galaxies as small semi-transparent disks with
radii equal to the effective radii of their corresponding Voronoi
cells. 

The user can directly compare two void catalogs by using a built-in 
function for computing the overlap. This function attempts to 
find a one-to-one match between voids in one catalog and another 
and reports the relative properties (size, ellipticity, etc.) 
between them. It also records which voids cannot find a reliable match 
and how many potential matches a void may have. 
A more detailed discussion of the matching process can be 
found in~\citet{Sutter2013b}.

\vide~allows the user to compute simple two-point clustering statistics, i.e. power spectra and correlation functions. 

Three-dimensional stacks of voids are useful in a variety of 
scenarios, and the user may construct these with built-in functions.
The stacked void 
may optionally contain only void member particles or all particles 
within a certain radius. 
With this stacked profile \vide~can build a 
spherically averaged radial density profile and fit
the universal HSW void profile to it. All proper 
normalizations are handled internally.
Figure~\ref{fig:numberfunc} shows an example of \vide-produced 
density profiles and best-fit HSW profile curves. 

\section{Applications}
\label{sec:applications}

Throughout its development \vide~has been used in a variety of applications.
Initially, the enhancements to \textsc{zobov} were used to create a set 
of public catalogs from both observations~\citep{Sutter2012a, Sutter2013c}
and simulations~\citep{Sutter2013a}. The toolkit presented in this work 
is fully compatible with those public releases.

\begin{figure} 
  \centering 
  {\includegraphics[type=png,ext=.png,read=.png,width=0.49\columnwidth]{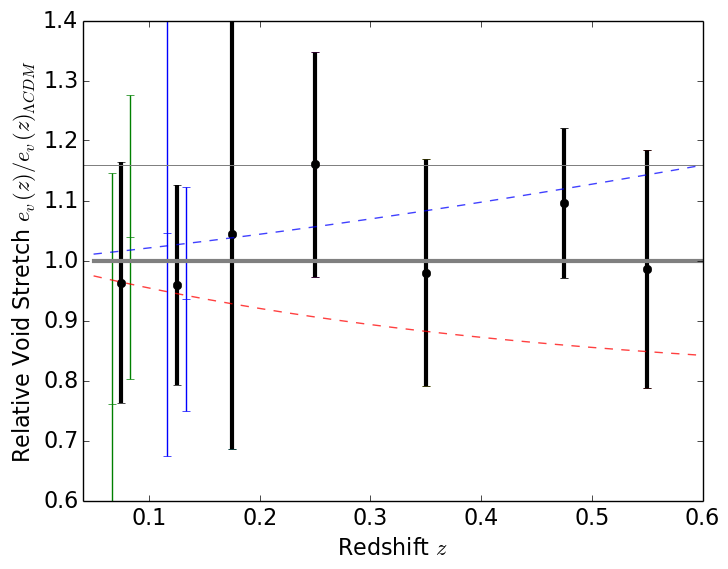}}
  {\includegraphics[type=png,ext=.png,read=.png,width=0.49\columnwidth]{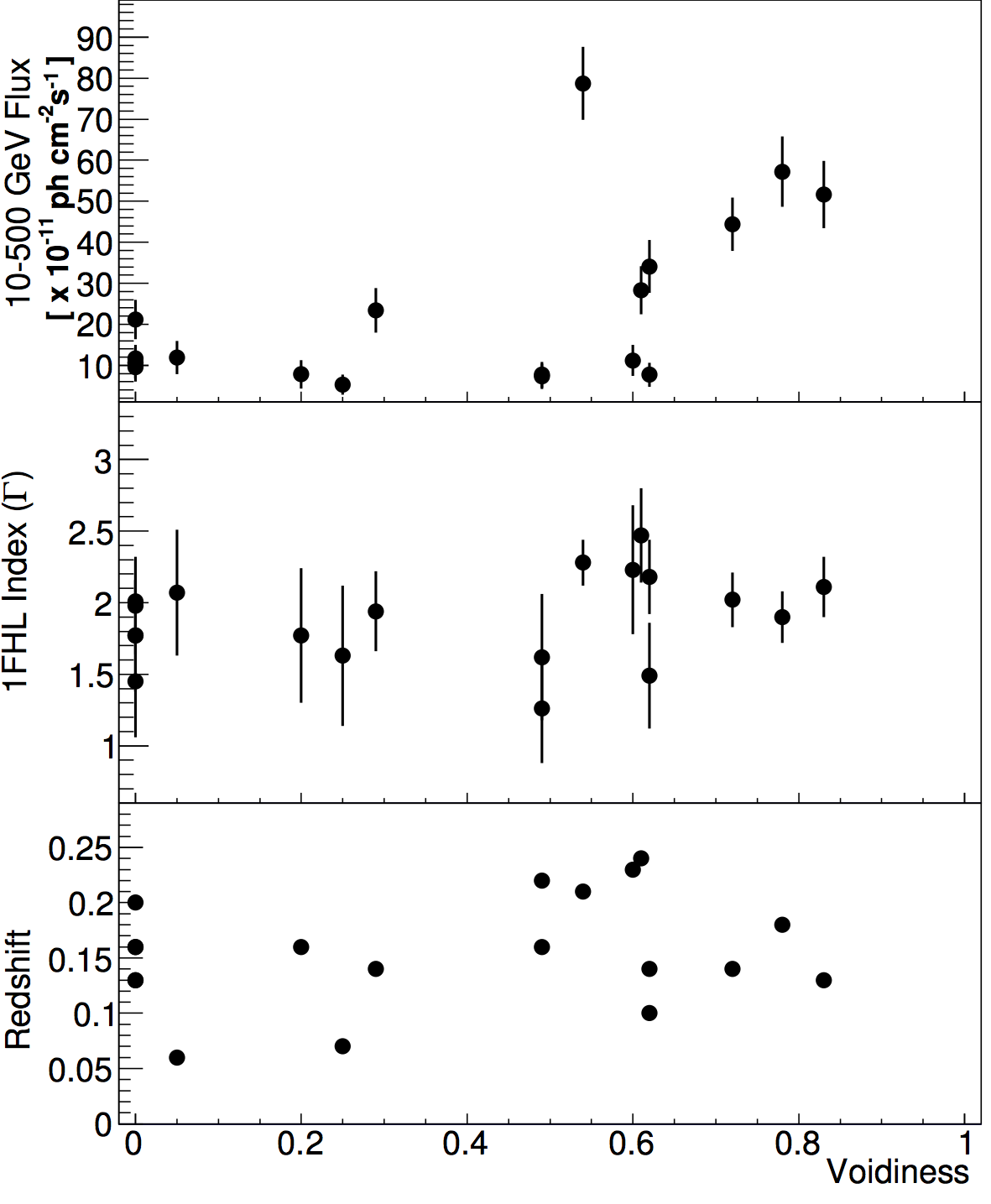}}
  \caption{ 
           Left panel: a detection of the \ap~effect using stacked cosmic voids 
           in the SDSS DR7 and DR10. The slight deviation from unity in the 
           expected void stretch can be used to extract cosmological 
           parameters. 
           \emph{Reproduced from~\citet{Sutter2014b}}.
           Right panel: GeV luminosity versus void fraction along the line of sight for Fermi-detected blazars.
           \emph{Reproduced from~\citet{Furniss2014}}.
           }
\label{fig:apps}
\end{figure}

The publication of simple void properties, such as sky position and 
effective radius, enabled direct cross correlations with other datasets
leading to several interesting results. Examples of such 
correlations include: 
galaxy shear maps leading to a detection of a weak lensing effect 
inside voids~\citep{Melchior2014}, 
the cosmic microwave background giving rise to a detection of the 
integrated Sachs-Wolfe effect~\citep{Planck2013b, Ilic2013}, and 
positions of high-energy blazars putting limits on the intergalactic
magnetic field~\citep{Furniss2014}. The latter correlation is shown 
in the right panel of Figure~\ref{fig:apps}.
Additionally, cross-correlations of void centers with galaxy positions
 have recently been demonstrated
 to allow the construction of 
a static ruler~\citep{Hamaus2013}, and the three-dimensional 
arrangement of galaxies around
stacked voids has been exploited for an \ap~
test~\citep{Sutter2014b}, as shown in the left-hand panel of 
Figure~\ref{fig:apps}.

\begin{figure} 
  \centering 
  {\includegraphics[type=png,ext=.png,read=.png,width=0.49\columnwidth]{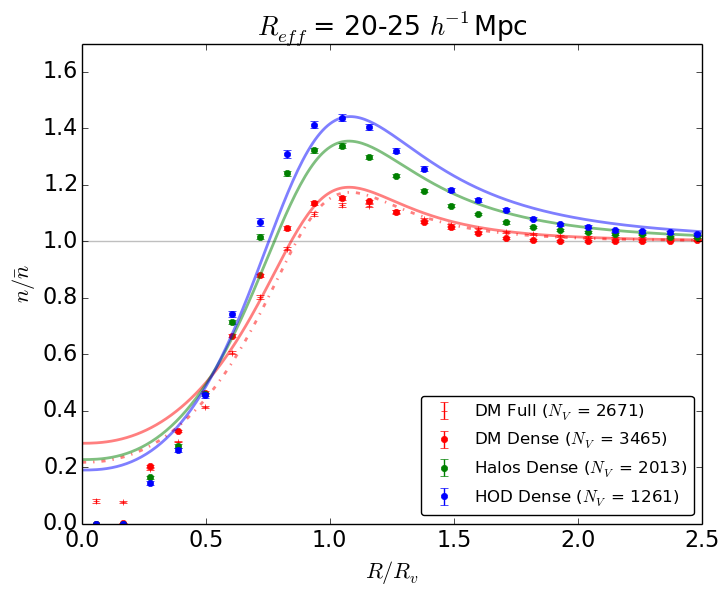}}
  {\includegraphics[type=png,ext=.png,read=.png,width=0.49\columnwidth]{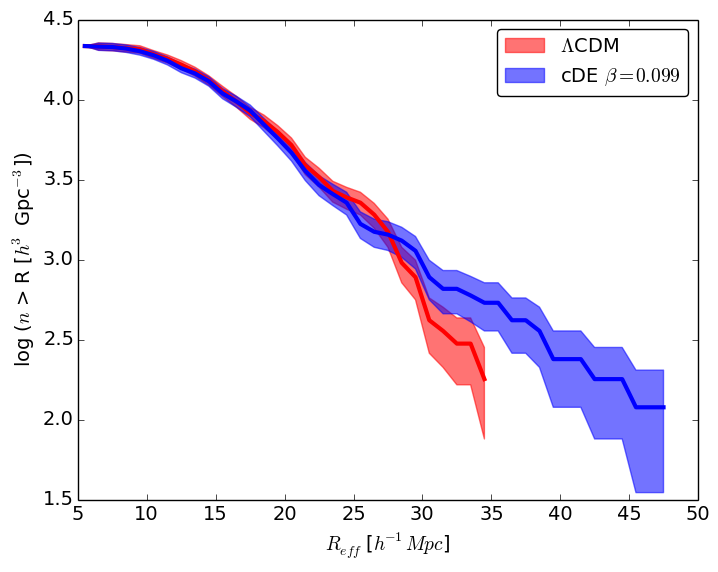}}
  \caption{ 
           Left panel: example one-dimensional radial density profiles of stacked voids
           (points with error bars) and best-fit curves (thin lines)
           using the HSW profile.
           \emph{Reproduced from~\citet{Sutter2013a}}.
           Right panel: example cumulative void number functions from simulations.
           \emph{Reproduced from~\citet{Sutter2014c}}.
           }
\label{fig:numberfunc}
\end{figure}

Derived void statistical information,
such as radial density profiles (as shown in Figure~\ref{fig:numberfunc}), have 
led to the discovery of a universal density 
profile~\citep{Hamaus2014,Sutter2013a} and a method for constructing 
real-space profiles in a parameter-independent manner~\citep{Pisani2013}.
The abundance of voids is potentially a very powerful probe of 
cosmology, as shown in~\citet{Sutter2014c} and highlighted
in Figure~\ref{fig:numberfunc}.

Lastly, the ability to match and compare void properties from one catalog 
to another (a built-in feature of \vide) has been used to examine the 
relationship of voids identified in sparse, biased galaxy samples to 
dark matter underdensities~\citep{Sutter2013a} and to reconstruct 
the life history of voids using merger trees~\citep{Sutter2014a}.

\section{Conclusions}
\label{sec:conclusions}

We have presented and discussed the capabilities of \vide, a new 
Python/C++ toolkit for identifying cosmic voids in cosmological 
simulations and galaxy redshift surveys, as well as a brief overview 
of its current applications. \vide~performs void 
identification using a substantially modified and enhanced version 
of the watershed code \textsc{zobov}. 
Furthermore, \vide~is able to support a variety of mock and real datasets,
and provides extensive and flexible tools for 
loading and analyzing void catalogs. We have highlighted these 
analysis tools (e.g., filtering, plotting, clustering statistics, stacking, 
profile fitting) using examples from previous and current 
research using \vide.

The analysis toolkit bundled in \vide~enables a wide variety of 
both theoretical 
and observational void research. 
As its most basic, the void properties made available to the 
user, such as sky positions, shapes, and sizes, permit simple explorations 
of void properties and cross-correlation with other datasets.
The user may also use void member particles and their associated 
volumes for examining galaxy properties in low-density 
environments. Cross-matching by overlapping catalogs can be
 useful for understanding the 
impacts of peculiar velocities or galaxy bias, as well as providing 
a platform for studying the effects of modified gravity or 
fifth forces on a void-by-void basis. 
Void power spectra, 
shape distributions, number functions, and density profiles, 
easily accessible via \vide, 
are sensitive probes of cosmology. Users may also access 
HSW density profiles, enabling theoretical predictions
of the ISW or gravitational lensing signals. 

The past few years have seen an enormous growth in void 
interest and research. This research includes new void-finding 
algorithms, studies of void properties, investigations and forecasts 
of cosmological 
probes, and explorations into the nature of void themselves from theoretical 
and numerical viewpoints. 
Put simply, \vide~is designed to meet the growing demand of next-generation 
void science.

The \vide~code and documentation is currently hosted at\linebreak {\tt http://bitbucket.org/cosmicvoids/vide\_public}, with links to numbered versions at\linebreak {\tt http://www.cosmicvoids.net}.

\section*{Acknowledgments}
The authors acknowledge
support from NSF Grant NSF AST 09-08693 ARRA. BDW
acknowledges funding from an ANR Chaire d'Excellence (ANR-10-CEXC-004-01),
the UPMC Chaire Internationale in Theoretical Cosmology, and NSF grants AST-0908
902 and AST-0708849.
This work made in the ILP LABEX (under reference ANR-10-LABX-63) was supported by French state funds managed by the ANR within the Investissements d'Avenir programme under reference ANR-11-IDEX-0004-02.

\bibliographystyle{mn2e}
\bibliography{sutteriau}

\end{document}